\documentclass[aps,prd,preprintnumbers,nofootinbib,preprint]{revtex4}
\usepackage{bm}
\usepackage{latexsym}
\usepackage{dcolumn}
\usepackage{amsmath,amsfonts,amssymb}
\usepackage{graphicx,epsfig}
\usepackage{amsthm}

\newcommand{\ve}{\delta}
\newcommand{\be}{\begin{eqnarray}}
\newcommand{\ee}{\end{eqnarray}}
\newcommand{\bea}{\begin{eqnarray}}
\newcommand{\eea}{\end{eqnarray}}

\def\comment#1{}

\newcommand{\lp}{\ell_{\rm p}}
\newcommand{\mpl}{m_{\rm p}}

\newcommand{\rh}{r_{\rm H}}

\newcommand{\ep}{\mathcal{E}_{\rm p}}

\usepackage{color}

\definecolor{darkred}{rgb}{.8,0,0}

\definecolor{darkblue}{rgb}{0,0,.7}

\definecolor{darkgreen}{rgb}{0,.7,0}

\begin{document}

%
%
\title{GUP parameter and black hole temperature} 
%
%
%
%
%
\author{{Elias~C.~Vagenas}$^1$}\email[email:~]{elias.vagenas@ku.edu.kw}
\author{Salwa~M.~Alsaleh$^2$}\email[email:~]{salwams@ksu.edu.sa}
\author{Ahmed~Farag~ Ali$^{3,4}$} \email[email:~]{ahmed.ali@fsc.bu.edu.eg}
\affiliation{$^1$Theoretical Physics Group, Department of Physics, Kuwait University, P.O. Box 5969, Safat 13060, Kuwait}
\affiliation{$^2$Department of Physics and Astronomy, King Saud University, Riyadh 11451, Saudi Arabia}
\affiliation{$^3$Netherlands Institute for Advanced Study, Korte Spinhuissteeg 3, 1012 CG Amsterdam, The Netherlands}
\affiliation{$^4$Department of Physics, Faculty of Science, Benha University, Benha, 13518, Egypt}
%
%
%
%
%
\begin{abstract}
%
\par\noindent
Motivated by a recent work of Scardigli, Lambiase and Vagenas (SLV), we derive the GUP parameter, i.e. $\alpha_0$, 
when the GUP has a linear 
and quadratic term in momentum. The value of the GUP parameter is obtained by conjecturing that the GUP-deformed 
black hole temperature of a Schwarzschild black hole and the modified Hawking temperature 
of a quantum-corrected Schwarzschild black hole are the same. The leading term in both cases 
is the standard Hawking temperature and since the corrections are considered as thermal,  the modified 
and deformed expressions of temperature display a slight shift in the Hawking temperature. Finally, by equating the 
first correction terms, we obtain a value for the GUP parameter. In our analysis, the GUP parameter is not a pure number but depends on the ratio $\mpl /M$ with $\mpl$ to be the Planck mass and $M$ the black hole mass. 
\end{abstract}

\maketitle
%
\section{Introduction}
%
%
%
\par\noindent
Gravitation, described by the relation between space-time curvature and the presence of stress-energy 
in General Relativity (GR), is incompatible with the  Heisenberg uncertainty principle (HUP), as the latter 
predicts that measurement of position to high accuracy would introduce a huge amount of energy to the system, 
thus, causing the space-time structure to break down due to gravity. Therefore, the HUP should be generalized to 
incorporate quantum gravitational effects.  Such generalizations were studied extensively in the past decade 
\cite{GUPearly}. Moreover, they were suggested by several candidate theories of quantum gravity, such as string theory, 
loop quantum gravity, deformed special relativity, and studies of black hole physics 
\cite{VenezGrossMende,MM,kempf,FS,Adler2,SC}.
\par\noindent
There are many possible deformations of the HUP to a generalized uncertainty principle (GUP) by a dimensionless 
deformation parameter $\alpha_0$. This parameter can be calculated on a theoretical basis (as in some models of string theory, 
see Ref.~\cite{VenezGrossMende}), or from a phenomenological approach that would set a bound on this deformation 
parameter, for example see Refs. \cite{brau,vagenas,Nozari}.
In most of the studies above, a  specific non-linear  representation of the operators in the deformed canonical 
commutation relations (CCR) is used 
\be
\left[\hat{X},\hat{P}\right] = i\,\hbar\left(1 + \alpha_{0}^{2}\, \frac{\hat{P}^{2}}{\mpl^{2}}\right)
\label{[1]}
\ee
\par\noindent
where $\alpha_0$ is the dimensionless GUP parameter and  $\mpl$ is the Planck mass
\footnote{Here we set $c=k_B=1$, thus the Planck length is given as $\lp^2=G\,\hbar$ while  the Planck energy 
will be $\ep =\mpl$. Moreover the Planck energy satisfies the HUP, i.e.,   $\ep\,\lp = \hbar /2$ ,and thus, 
the Planck mass and Planck length are connected through $\hbar=2\,\lp\,\mpl$.}.
\par\noindent
However, a more general deformation was explored in \cite{Ali}, that has a linear and a quadratic term in momentum $P$
\begin{equation}
[\hat X,\hat P] = i \hbar \left( 1- 2\alpha \hat P +  4\alpha^2 \hat P^2 \right)
\label{communator}
\end{equation}
%
%
%
\par\noindent
with $\alpha = \alpha_{0}/ \mpl = 2\alpha_{0}\lp / \hbar$.

\par\noindent
The deformed representations of the Heisenberg algebra of the CCR lead to deformed 
quantum mechanics, which leads to deducing discreteness of space-time as in Ref. \cite{Ali}. 
Moreover, the deformed CCR were used in calculations of corrected energy shift of the hydrogen atom 
spectrum, the Lamb shift, the Landau levels, and  the Scanning Tunneling Microscope (STM), 
all leading to estimates on $ \alpha_0$ ranging between $\alpha_0 < 10^{11}$ and $\alpha_0 < 10^{25}$ 
\cite{vagenas,Das:2009hs}.
\par\noindent
A more refined approach for evaluating bounds on $ \alpha_0$ from gravitational interaction was made 
in Ref.~\cite{SC2}. In this work, the Poisson brackets and classical Newtonian mechanics were not 
deformed, unlike the previous approaches. Moreover, the authors were able to recover standard GR and quantum 
mechanics when $ \alpha_0 \to 0$. Hence, general covariance is preserved, and there is no effect on the geodesic 
equation for a test particle in the gravitational field. The bounds on $\alpha_0$ obtained from the gravitational interaction 
range between  $\alpha_{0}<10^{10}$ and  $\alpha_{0} < 10^{35}$.
\par
However, it is believed that $ \alpha_0$ should be of order unity, as predicted by string 
theory \cite{VenezGrossMende}. Closer bounds to this were obtained from considering the proton decay by GUP-deformed 
virtual black holes \cite{alsaleh} which obtained a bound of $ \alpha_{0} > 10^{-3}$, and another closer to 
unity $ \alpha_{0} \sim 1$ when higher dimensions are considered. 
\par
In a previous  work \cite{Scardiglia},  a computation of the value of $\alpha_0$ was performed by comparing
two different low energy (first order in $\hbar$) corrections for the expression of the Hawking temperature.
The first is due to a  quadratic GUP, and therefore involves $\beta_0$ which is equivalent to $\alpha_{0}^2$. 
The second correction  is obtained by including the deformation of the metric due to quantum corrections to the Newtonian potential. 
The value obtained was also of order of unity $ \alpha_{0}^2 =  \displaystyle\frac{82\pi}{5}$, 
in agreement with the predictions of string theory. 
\par
In this paper, we extend these computations to a linear and quadratic GUP, 
obtaining a GUP parameter $\alpha_0$ which depends on the ratio 
$\mpl /M$ with $\mpl$ to be the Planck mass and $M$ the black hole mass. 
%
%
%
%
\section{GUP-deformed Black hole temperature}
%
%
\par\noindent
In this section, we use a very general deformation of the HUP \cite{Ali}, with linear and quadratic terms in momentum $p$
\begin{equation}
\Delta x \Delta p \geq  \frac{\hbar}{2}\left[ 1+ \left( \frac{\alpha}{\sqrt{\langle p^2 \rangle}}+4 \alpha^2\right) 
\Delta p^2 + 4 \alpha^2 \langle p\rangle ^2 -2 \alpha \sqrt{\langle p^2 \rangle}\right] ~.
\label{GUP}
\end{equation}
\par\noindent
Since we are interested in mirror-symmetric states, such that $ \langle p\rangle ^2 =0$  , hence 
$\Delta p = \sqrt{\langle p^2 \rangle}$, and  $\Delta x\, \Delta p \geq (1/2) |\langle [\hat{x},\hat{p}] \rangle|$, 
the above-mentioned GUP  in terms of commutators reads
\begin{equation}
[x,p] = i \hbar \left( 1- 2\alpha p +  4\alpha^2 p^2 \right)~.
\label{communator1}
\end{equation}
\par\noindent
Based on the Heisenberg microscope argument \cite{Heisenberg}, if one wants to locate a particle of 
size $ \delta x$, then a photon of energy $E$ has to be shot towards this particle. Employing the GUP version 
given by equation  \eqref{GUP} and following the arguments in 
Refs.~\cite{FS9506,ACSantiago,CavagliaD,CDM03,Susskind,nouicer,Glimpses}, the size of the particle will roughly be 
\footnote{The standard dispersion relation $E=p$ holds true.}
\begin{equation}
\delta x \sim \frac{\hbar }{2 E}- \frac{\hbar \alpha}{2} + 2 \hbar \alpha^2 E ~.
\label{heisenberg}
\end{equation}
\par\noindent 
Of course, given the  particle's  (average) wavelength $\lambda\simeq \delta x$, 
one can use equation  (\ref{heisenberg}) to compute the energy $E$ of the particle. 
Following this syllogism, one can compute the GUP-deformed black hole temperature. 
In particular, one considers a Schwarzschild black hole of mass $M$ and event horizon at position $\rh = 2M G$. 
An ensemble of unpolarized photons which are the Hawking radiation particles, are coming out of this event horizon. 
The position uncertainty $\delta x$ of these photons is proportional to the size of the event horizon, namely 
$\delta x = 2 \mu \rh$, with $\mu$ to be a dimensionless constant which will be determined later. Based on the 
equipartition principle, the energy $E$ of the photons of the Hawking radiation is actually the temperature $T$ of the 
Schwarzschild black hole, and, thus, in this case  equation (\ref{heisenberg})  now reads
\begin{equation}
4 \mu G M \simeq \frac{ \hbar}{2 T}   -  \frac{\hbar \alpha}{2} + 2\hbar  \alpha^{2} T 
\label{temp}
\end{equation}
\par\noindent
or, equivalently, 
\begin{equation}
4 \mu G M \simeq \frac{ \hbar}{2 T}   -  \frac{\hbar \alpha_{0}}{2\mpl} + \frac{2\hbar  \alpha_{0}^{2}}{\mpl^{2}} T ~.
\label{temp1}
\end{equation}
\par\noindent
At this point, we can fix the parameter $ \mu$ by taking the deformation parameter $ \alpha_0 \to 0$ and 
requiring to recover the standard Hawking temperature, i.e.,
\begin{equation}
T_{BH} = \frac{\hbar}{8 \pi G M}.
\label{tbh}
\end{equation}
Therefore, the dimensionless constant has to  be $ \mu = \pi$.  Next, we rearrange equation \eqref{temp1} 
in order to make it a quadratic equation of the black hole temperature $T$ 
\begin{equation}
\left(\frac{4\hbar\alpha_0^2}{\mpl^{2}}\right) T^2 - \left(8\pi G M + \frac{\hbar\alpha_0}{\mpl}\right) T +\hbar =0
\end{equation}
with roots
\begin{widetext}
\begin{equation}
T = \frac{\left(8\pi G M + \displaystyle\frac{\hbar\alpha_0}{\mpl}\right)\pm\sqrt{\left(8\pi G M +
\displaystyle \frac{\hbar\alpha_0}{\mpl}\right)^{2} 
-4\left(\displaystyle\frac{4\hbar\alpha_0^2}{\mpl^{2}}\right) \hbar}}
{2\left(\displaystyle\frac{4\hbar\alpha_0^2}{\mpl^{2}}\right)}~.
\label{T}
\end{equation}
\end{widetext}
Now, we can expand equation  \eqref{T}  near $\alpha_0 \to 0$ and up to second order in $\alpha_{0}$, 
and, thus,  we obtain
\begin{equation}
T =  \frac{\hbar }{8 \pi  G M} \left[  1 - \frac{\alpha_0}{2\pi}\left(\frac{\mpl}{M}\right) + 
5\left(\frac{ \alpha_0  }{2\pi }\right)^{2}\left(\frac{\mpl}{ M}\right)^{2}\right] .
\label{tgup}
\end{equation}
\par\noindent
At this point it should be stressed that we assumed that the GUP corrections have a thermal character.  
Thus, it was expected that  the GUP corrections will  produce a  slight shift in the Hawking spectrum, 
and, therefore, the GUP-deformed black hole temperature is a shifted Hawking temperature.
%
%
%
%
\section{Quantum-corrected Schwarzschild metric}
%
%
%
\par\noindent
General relativity can be considered as an effective field theory, when considering two heavy objects 
at rest. This leads to tree diagrams for graviton exchange between these objects as first studied by Duff \cite{Duff:1974ud}. 
The quantum correction to the Newton's potential from these diagrams was obtained (up to a leading term) by Donoghue 
\cite{Dono}. Furthermore, it was found that the gravitational interaction between the two above-mentioned 
objects 
%
%
%
%
%
%
%
%
 %
 %
can be described by a potential energy which is produced by the potential generated from the mass $M$  \cite{Dono2}
 \be
 V(r)=-\frac{GM}{r}\left(1 + \frac{3GM}{r}(1+\frac{m}{M}) +  \frac{41}{10\pi}\frac{\ell_P^2}{r^2}\right) .
 \label{DP}
 \ee
\par\noindent
Since the quantum corrections were made to the Schwarzschild solution, we expect, as mentioned in 
\cite{Duff:1974ud}, that the classical Schwarzschild metric 
\be
ds^2 = \left(1-\frac{2GM}{r}\right)dt^2 - \left(1-\frac{2GM}{r}\right)^{-1}dr^2 - r^2d\Omega^2
\ee
would be deformed. This can be found easily from the weak-field limit approximation, considering a large mass 
black hole, e.g., a  solar-mass black hole. We have the relation
\be
V(r) \ \simeq \ \frac{1}{2} \, (g_{tt}(r)-1)
\ee
or,  equivalently,
\be
g_{tt}(r) \ \simeq \ 1 \ + \ 2 \ V(r)~.
\ee
\par\noindent
Thus, we can find the $tt$ component of the deformed metric
\be
g_{tt}(r) = 1 - \frac{2GM}{r} + \epsilon(r)~.
\label{DM}
\ee
Here we have introduced $ \epsilon(r)$  
%
%
%
\be
\epsilon(r) = - \frac{6\, G^2 M^2}{r^2}\left(1+\frac{m}{M}\right) -
\frac{41}{5\pi}\frac{G^3 M^3}{r^3}\left(\frac{\ell_P}{GM}\right)^{2}\hspace{-1.5ex}.~
\label{phi}
\ee
\par\noindent
It can be easily seen that the  general form of the Schwarzschild metric can be written as 
\cite{Weinberg72}
\be
ds^2 = F(r)dt^2 - F(r)^{-1}dr^2 - C(r)d\Omega^2
\label{gm}
\ee
where $ F(r)= g_{tt}$.
The effective Newtonian potential can be derived from a deformed metric  for  a point particle that moves at non-relativistic 
velocities, in a stationary gravitational field, i.e.,  asymptotically quasi-Minkowskian  $r \to \infty$. 
Hence, we obtain the quantum-corrected metric 
\begin{widetext}
\be
ds^2 = \left(1 - \frac{2GM}{r} + \epsilon(r)\right) dt^2 - \left( 1 - \frac{2GM}{r} + \epsilon(r)\right) ^{-1}dr^{2} - 
r^2d\Omega^2 ~.
\label{qsch}
\ee
\end{widetext}
%
%
%
%
\section{Computing the GUP parameter}
%
%
%
%
\par\noindent
In this section, we follow the methodology developed in Ref. \cite{Scardiglia} in order to 
compute the quantum-corrected Hawking temperature from the metric given by equation \eqref{qsch}. 
We employ the general formula for the black hole temperature \cite{Angheben:2005rm}
\be
T=  \frac{\hbar}{4 \pi}\,\lim_{r\to\rh}\left[ g_{tt}(r) \right]'
\ee
\par\noindent
with the prime `` $' $ " to denote the partial derivative with respect to the radial coordinate, i.e., $r$, 
and $\rh$ to be  the solution of the horizon equation
\be
r - 2 \,G\, M + \epsilon(r)\, r = 0~.
\label{eqr}
\ee
\par\noindent
It should be noted that this deformation is valid in the limit $|\epsilon(r)| \ll GM/r$. Next, we  ``regularize"
$\epsilon(r)$ by writing $ \epsilon(r) \equiv \delta \phi(r)$ with $ \phi(r)$ being a smooth function of $r$, 
and $\delta$  a regularization parameter. 
Hence, the solution to equation \eqref{eqr} reads \cite{Scardiglia} 
\be
\rh \ = \ a - \frac{\ve\, a\, \phi(a)}{1 \ + \ \ve\,[\phi(a) \ + \ a\,\phi'(a)]}
\label{solr}
\ee
where  $a = 2 G M$. 
Moreover, the value of $\left[ g_{tt}(\rh)\right]'$ is equal to
\be
\left[ g_{tt}(\rh)\right]' = \frac{1}{a(1-\lambda)^2} \ + \ \ve \phi'[a(1-\lambda)]
\ee
\par\noindent
with the parameter $\lambda$ to be 
\be
\lambda =   \frac{\ve\, \phi(a)}{1 \ + \ \ve\,[\phi(a) \ + \ a\,\phi'(a)]}~.
\ee
\par\noindent
Therefore, the quantum-corrected Hawking temperature of the Schwarzschild black hole is given by
 \be
 T &=& \, \frac{\hbar}{4\pi}\,g_{tt}'(\rh) = \frac{\hbar}{4\pi a} \left\{1 + \ve \left[2\phi(a) + a\phi'(a)\right] \right. \nonumber \\
 &+& \left. \ve^2 \phi(a) \left[\phi(a) - 2a\phi'(a) - a^2 \phi''(a) \right] + \dots \right\}~.
 \label{teps}
 \ee
\par\noindent
At this point, we restore $\epsilon(r)$  and take only the first order terms in $\delta$, thus
\begin{equation}
T = \frac{\hbar}{8 \pi G M} \Big (1+ [ 2 \epsilon(a)+ a \epsilon'(a) ]  \Big) ~.
\label{Tq}
\end{equation}
\par\noindent
Now, we may conjecture that the GUP-deformed black hole temperature given by equation \eqref{tgup} 
is equal to the quantum-corrected Hawking temperature given by equation \eqref{Tq}. Thus, the GUP deformation 
corresponds to the tree-diagram quantum correction to gravity. 
Using this conjecture, we can compare equation \eqref{tgup} with equation \eqref{Tq} to obtain an estimate for the GUP 
parameter $ \alpha_0$. Utilizing equation (\ref{phi}), for the RHS of  equation \eqref{Tq}, we obtain
\begin{equation}
2 \epsilon(a)+ a \epsilon'(a) = \frac{B}{8 G^3 M^3}
\end{equation}
\par\noindent
and $ B=\displaystyle\frac{41 G^2 M \hbar }{5 \pi }$.
\par\noindent
We can now obtain a value for $ \alpha_0$ by comparing the first and second order terms in $\alpha_0$ 
of equation \eqref{tgup}  with the first order corrections in $\epsilon$ of equation \eqref{Tq}, i.e.,
\begin{equation}
- \frac{\alpha_0}{2\pi}\left(\frac{\mpl}{M}\right)  
+ 5\left(\frac{\alpha_0  }{2\pi}\right)^{2}\left(\frac{\mpl}{ M}\right)^{2}
=\frac{41}{10 \pi }\left(\frac{\mpl}{M}\right)^{2}~.
\end{equation}
\par\noindent
The solutions of  this quadratic equation read
\begin{equation}
\alpha_0 =\frac{ \displaystyle\left(\frac{1}{2\pi}\left(\frac{\mpl}{M}\right)\right) \pm  
\sqrt{\displaystyle\left(\frac{1}{2\pi}\left(\frac{\mpl}{M}\right)\right)^{2}+
4\left(\frac{41}{10\pi}\right)\left(\frac{5}{4\pi^2}\right)}}
{\displaystyle 2\left(\frac{5}{4\pi^2}\right)}~.
\label{quada}
\end{equation}
It is noteworthy that equation  \eqref{quada} depends on $ M$. Since $ M >> m_p$,  
we may expand the negative solution in expression (\ref{quada}) and 
the leading terms  give the value for $ \alpha_0 $
\begin{equation}
\alpha_0 = - \frac{82}{10} \left( \frac{\mpl}{M}\right)+ \frac{1681}{10\pi}\left( \frac{\mpl}{M}\right)^{3}~.
\label{alpha}
\end{equation}
%
%
%
%
\section{Conclusions}
%
%
%
%
%
\par\noindent
In this paper, we have calculated the value of the deformation parameter  $\alpha_0$ of a linear and quadratic GUP. 
We obtained this value by conjecturing that the deformed Hawking temperature  due to GUP 
is equivalent to the Hawking temperature obtained from a quantum-corrected Schwarzschild solution 
when considering a tree diagram between two massive objects gravitationally interacting.
\par
The GUP-deformed Hawking temperature was obtained by assuming a particle localized near the horizon,  
$ \Delta x \sim \rh$, and using a GUP with linear and quadratic terms in $p$ instead of the standard HUP. 
In this way, we obtained an expression for the black hole temperature with terms depending on 
the GUP deformation parameter $ \alpha_0$ (see equation (\ref{tgup})).
\par
The quantum corrections of the Schwarzschild black hole metric stem from the quantum correction to the Newtonian 
potential derived  by Donoghue and collaborators. Specifically, the corrections to the Newtonian potential imply naturally 
a quantum correction to the Schwarzschild metric, in the weak-field limit . 
Using this corrected metric, we were able to obtain an expression for the quantum-corrected temperature 
(see equation (\ref{Tq})).
\par
Finally, we  compared both expressions for the Hawking temperature and equated the subleading terms, 
since the leading terms are the same and equal to the standard Hawking temperature. 
As we conjectured that the GUP-deformed black hole temperature is equivalent to the quantum-corrected temperature, 
we have obtained an expression for $\alpha_0$.
\par
This work is an extension of the work done in Ref.~\cite{Scardiglia} in the sense that in Ref.~\cite{Scardiglia}  
a GUP with a quadratic term in momentum is used while in this work we have employed a GUP with linear and 
quadratic terms in momentum.
However,  the GUP parameter $\alpha_{0}$  obtained in this work is proportional to powers of the dimensionless
ratio $(\mpl / M)$ and this means that the GUP parameter $\alpha_{0}$  depends on the specific system under study. 
Therefore, this result can be interpreted as a loss  of or ``less"  universality of GUP compared to the one   described in Ref.~\cite{Scardiglia}. 
%
%
%
%
%
%
%
%
%
%
%
%
%

%
%
%
\end{document}